\documentstyle[12pt,epsfig]{article}

\catcode`\@=11
\def\@citex[#1]#2{%
\if@filesw \immediate \write \@auxout {\string \citation {#2}}\fi
\@tempcntb\m@ne \let\@h@ld\relax \def\@citea{}%
\@cite{%
  \@for \@citeb:=#2\do {%
    \@ifundefined {b@\@citeb}%
      {\@h@ld\@citea\@tempcntb\m@ne{\bf ?}%
      \@warning {Citation `\@citeb ' on page \thepage \space undefined}}%
      {\@tempcnta\@tempcntb \advance\@tempcnta\@ne%
      \@tempcntb\number\csname b@\@citeb \endcsname \relax%
      \ifnum\@tempcnta=\@tempcntb %Number follows previous--hold on to it
        \ifx\@h@ld\relax%
          \edef \@h@ld{\@citea\csname b@\@citeb\endcsname}%
        \else%
          \edef\@h@ld{\ifmmode{-}\else--\fi\csname b@\@citeb\endcsname}%
        \fi%
      \else%   %  non-successor--dump what's held and do this one
        \@h@ld\@citea\csname b@\@citeb \endcsname%
        \let\@h@ld\relax%
      \fi}%
    \def\@citea{,\penalty\@highpenalty\,}%
  }\@h@ld
}{#1}}

\def\@citeb#1#2{{[#1]\if@tempswa , #2\fi}}
\def\@citeu#1#2{{$^{#1}$\if@tempswa , #2\fi }}
\def\@citep#1#2{{#1\if@tempswa , #2\fi}}

\def\bcites{         % cite with []'s
        \catcode`\@=11
        \let\@cite=\@citeb
        \catcode`\@=12
}

\def\upcites{         % cite with exponents
        \catcode`\@=11
        \let\@cite=\@citeu
        \catcode`\@=12
}

\def\plaincites{      % cite without brackets
        \catcode`\@=11
        \let\@cite=\@citep
        \catcode`\@=12
}

\newcount\hour
\newcount\minute
\newtoks\amorpm
\hour=\time\divide\hour by 60
\minute=\time{\multiply\hour by 60 \global\advance\minute by-\hour}
\edef\standardtime{{\ifnum\hour<12 \global\amorpm={am}%
        \else\global\amorpm={pm}\advance\hour by-12 \fi
        \ifnum\hour=0 \hour=12 \fi
        \number\hour:\ifnum\minute<10 0\fi\number\minute\the\amorpm}}
\edef\militarytime{\number\hour:\ifnum\minute<10 0\fi\number\minute}

\def\draftlabel#1{{\@bsphack\if@filesw {\let\thepage\relax
   \xdef\@gtempa{\write\@auxout{\string
      \newlabel{#1}{{\@currentlabel}{\thepage}}}}}\@gtempa
   \if@nobreak \ifvmode\nobreak\fi\fi\fi\@esphack}
        \gdef\@eqnlabel{#1}}
\def\@eqnlabel{}
\def\@vacuum{}
\def\marginnote#1{}
\def\draftmarginnote#1{\marginpar{\raggedright\scriptsize\tt#1}}
\def\draft{
        \pagestyle{plain}
        \overfullrule=2pt
        \oddsidemargin -.5truein
        \def\@oddhead{\sl \phantom{\today\quad\militarytime} \hfil
        \smash{\Large\sl DRAFT} \hfil \today\quad\militarytime}
        \let\@evenhead\@oddhead
        \let\label=\draftlabel
        \let\marginnote=\draftmarginnote
        \def\ps@empty{\let\@mkboth\@gobbletwo
        \def\@oddfoot{\hfil \smash{\Large\sl DRAFT} \hfil}
        \let\@evenfoot\@oddhead}
        \def\@eqnnum{(\theequation)\rlap{\kern\marginparsep\tt\@eqnlabel}%
        \global\let\@eqnlabel\@vacuum}  }

\def\blackfonts{
        \font\blackboard=msbm10 scaled\magstep1
        \font\blackboards=msbm8
        \font\blackboardss=msbm6
}

\def\nblack{            % For people without blackboard fonts
        \def\ZZ{{Z \n{10} Z}}
        \def\NN{{N \n{14} N}}
        \def\CC{{C \n{11} C}}
        \def\RR{{R \n{11} R}}
        \def\QQ{{Q \n{12} Q}}
        \def\PP{{P \n{11} P}}
}

\def\prep{         % twocolumn.sty  Changed by Marek and Neil
        \catcode`\@=11
        \input art10.sty
        \catcode`\@=12
        
        \let\small\null
        \def\blackfonts{
                \font\blackboard=msbm10
                \font\blackboards=msbm7
                \font\blackboardss=msbm5
        }
        \let\sl\it
        \twocolumn
        \sloppy
        \voffset=-2.54truecm
        \hoffset=-2.54truecm
        \flushbottom
        \parindent 1em
        \leftmargini 2em
        \leftmarginv .5em
        \leftmarginvi .5em
        \marginparwidth 48pt
        \marginparsep 10pt
        \setlength{\columnsep}{2truecm}
        \setlength{\textwidth}{25.4truecm}
        \setlength{\textheight}{17truecm}
        \baselineskip=16pt
        \oddsidemargin .18truein
        \evensidemargin .17truein
}

\def\eqalign#1{\null\,\vcenter{\openup\jot\m@th
  \ialign{\strut\hfil$\displaystyle{##}$&$\displaystyle{{}##}$\hfil
      \crcr#1\crcr}}\,}
\def\eqalignno#1{\displ@y \tabskip\centering
  \halign to\displaywidth{\hfil$\@lign\displaystyle{##}$\tabskip\z@skip
    &$\@lign\displaystyle{{}##}$\hfil\tabskip\centering
    &\llap{$\@lign##$}\tabskip\z@skip\crcr
    #1\crcr}}

\def\section{\@startsection {section}{1}{\z@}{3.ex plus 1ex minus
 .2ex}{2.ex plus .2ex}{\large\bf}}
\def\subsection{\@startsection{subsection}{2}{\z@}{2.75ex plus 1ex minus
 .2ex}{1.5ex plus .2ex}{\bf}}        

\def\appendix{{\newpage\section*{Appendix}}\let\appendix\section%
        {\setcounter{section}{0}
        \gdef\thesection{\Alph{section}}}\section}

\def\abstract{\if@twocolumn
\section*{Abstract}
\else %\small
\begin{center}
{\bf Abstract\vspace{-.5em}\vspace{0pt}}
\end{center}
\quotation
\fi}

\catcode`\@=12

\newcommand{\beq}{\begin{equation}}
\newcommand{\eeq}{\end{equation}}
\newcommand{\beqa}{\begin{eqnarray}}
\newcommand{\eeqa}{\end{eqnarray}}

\newcommand{\tilQ}{\widetilde{Q}}

\newcommand{\dd}{{\rm d}}

\def\noj#1,#2,{{\bf #1} (19#2)\ }
\def\jou#1,#2,#3,{{\sl #1\/ }{\bf #2} (19#3)\ }
\def\ann#1,#2,{{\sl Ann.\ Physics\/ }{\bf #1} (19#2)\ }
\def\cmp#1,#2,{{\sl Comm.\ Math.\ Phys.\/ }{\bf #1} (19#2)\ }
\def\ma#1,#2,{{\sl Math.\ Ann.\/ }{\bf #1} (19#2)\ }
\def\ng#1,#2,{{\sl Nagoya.\ Math.\ J.\/ }{\bf #1} (19#2)\ }
\def\jd#1,#2,{{\sl J.\ Diff.\ Geom.\/ }{\bf #1} (19#2)\ }
\def\invm#1,#2,{{\sl Invent.\ Math.\/ }{\bf #1} (19#2)\ }
\def\cq#1,#2,{{\sl Class.\ Quantum Grav.\/ }{\bf #1} (19#2)\ }
\def\cqg#1,#2,{{\sl Class.\ Quantum Grav.\/ }{\bf #1} (19#2)\ }
\def\ijmp#1,#2,{{\sl Int.\ J.\ Mod.\ Phys.\/ }{\bf A#1} (19#2)\ }
\def\jmphy#1,#2,{{\sl J.\ Geom.\ Phys.\/ }{\bf #1} (19#2)\ }
\def\jams#1,#2,{{\sl J.\ Amer.\ Math.\ Soc.\/ }{\bf #1} (19#2)\ }
\def\grg#1,#2,{{\sl Gen.\ Rel.\ Grav.\/ }{\bf #1} (19#2)\ }
\def\mpl#1,#2,{{\sl Mod.\ Phys.\ Lett.\/ }{\bf A#1} (19#2)\ }
\def\nc#1,#2,{{\sl Nuovo Cim.\/ }{\bf #1} (19#2)\ }
\def\np#1,#2,{{\sl Nucl.\ Phys.\/ }{\bf B#1} (19#2)\ }
\def\pl#1,#2,{{\sl Phys.\ Lett.\/ }{\bf #1B} (19#2)\ }
\def\pla#1,#2,{{\sl Phys.\ Lett.\/ }{\bf #1A} (19#2)\ }
\def\pr#1,#2,{{\sl Phys.\ Rev.\/ }{\bf #1} (19#2)\ }
\def\prd#1,#2,{{\sl Phys.\ Rev.\/ }{\bf D#1} (19#2)\ }
\def\prl#1,#2,{{\sl Phys.\ Rev.\ Lett.\/ }{\bf #1} (19#2)\ }
\def\prp#1,#2,{{\sl Phys.\ Rept.\/ }{\bf #1C} (19#2)\ }
\def\ptp#1,#2,{{\sl Prog.\ Theor.\ Phys.\/ }{\bf #1} (19#2)\ }
\def\ptpsup#1,#2,{{\sl Prog.\ Theor.\ Phys.\/ Suppl.\/ }{\bf #1} (19#2)\ }
\def\rmp#1,#2,{{\sl Rev.\ Mod.\ Phys.\/ }{\bf #1} (19#2)\ }
\def\yadfiz#1,#2,#3[#4,#5]{{\sl Yad.\ Fiz.\/ }{\bf #1} (19#2) #3%
\ [{\sl Sov.\ J..\ Nucl.\ Phys.\/ }{\bf #4} (19#2) #5]}
\def\zh#1,#2,#3[#4,#5]{{\sl Zh.\ Exp.\ Theor.\ Fiz.\/ }{\bf #1} (19#2) #3%
\ [{\sl Sov.\ Phys.\ JETP\/ }{\bf #4} (19#2) #5]}

\def\beq{\begin{equation}}
\def\eeq{\end{equation}}
\def\beqar{\begin{eqnarray}}
\def\eeqar{\end{eqnarray}}

\newcommand{\be}{\begin{equation}}
\newcommand{\ee}{\end{equation}}
\newcommand{\bea}{\begin{eqnarray}}
\newcommand{\eea}{\end{eqnarray}}

\def\nfrac#1#2{{\displaystyle{\vphantom1\smash{\lower.5ex\hbox{\small$#1$}}%
        \over\vphantom1\smash{\raise.25ex\hbox{\small$#2$}}}}}

\def\n#1{\mskip-#1mu}

\def\to{\rightarrow}

\def\lae{\mathrel{\mathop{\smash{\lower .5 ex \hbox{$\stackrel<\sim$}}}}}
\def\lae{\mathrel{\mathop{\smash{\lower .5 ex \hbox{$\stackrel>\sim$}}}}}

\def\Tr{{\rm Tr}}
\def\l:{\mathopen{:}\,}
\def\r:{\,\mathclose{:}}

\catcode`\@=11
\def\theequation{\arabic{equation}}
\catcode`\@=12

\nblack
\bcites

\nblack

\catcode`\@=11
\def\theequation{\thesection.\arabic{equation}}
\@addtoreset{equation}{section}
\@addtoreset{footnote}{section}
\@addtoreset{footnote}{subsection}
\catcode`\@=12

\newcommand{\beqn}{\begin{equation}}
\newcommand{\eeqn}{\end{equation}}
\newcommand{\beqnarray}{\begin{eqnarray}}
\newcommand{\eeqnarray}{\end{eqnarray}}
\newcommand {\bear} [1] {\begin {array} {#1}}
\newcommand {\ear} {\end {array}}

\newcommand {\beqarn} {\begin{eqnarray*}}
\newcommand {\eeqarn} {\end{eqnarray*}}

\begin{document}

\begin{titlepage}

\begin{center}
\today
%Aug. xx, 1996
\hfill LBNL-41152, UCB-PTH-97/61, NSF-ITP-98-002\\
\hfill                  hep-th/9802005

\vskip 1.5 cm
{\large \bf Membrane Scattering in Curved Space\\ 
with M-Momentum Transfer}
\vskip 1 cm 
{Jan de\thinspace Boer$^{1,2}$, Kentaro Hori$^{1,2,3}$, 
and Hirosi Ooguri$^{1,2,3}$}\\
\vskip 0.5cm
{\sl $^1$Department of Physics,
University of California\\
Berkeley, CA 94720-7300, U.S.A.\\
~\\
$^2$Ernest Orlando Lawrence Berkeley National Laboratory\\
 Mail Stop 50A--5101, Berkeley, CA 94720, U.S.A.\\
~\\
$^3$Institute for Theoretical Physics, 
University of California\\ Santa Barbara, CA 93106, U.S.A\\}

\end{center}

\vskip 0.5 cm
\begin{abstract}
We study membrane scattering in a curved space with non-zero
M-momentum $p_{11}$ transfer. In the low-energy short-distance region,
the membrane dynamics is described by a three-dimensional $N=4$
supersymmetric gauge theory. We study an $n$-instanton process 
of the gauge theory, corresponding to the exchange of $n$ units of 
$p_{11}$, and compare the result with the scattering amplitude computed
in the low-energy long-distance region using supergravity. We find that
they behave differently.
%In particular, we show 
%that the instanton action of the gauge theory is less than
%the geodesic distance on the moduli space of vacua 
%and is larger than the Euclidean distance in the total field space. 
%When $n$ becomes large, 
%the difference of the instanton action and the geodesic 
%distance grows at least linearly in $n$. 
%and we discuss implications of this result to the large-$N$
%Matrix Theory conjecture. 
We show that this result is not in contradiction with the large-$N$
Matrix Theory conjecture,
by pointing out that cutoff scales of the supergravity 
and the gauge theory are complementary to each other.

\end{abstract}

\end{titlepage}

\section{Introduction}

There are three basic length scales in M Theory 
compactified on a circle
\cite{shenker}: 
the IIA string
length $l_s$, the Planck length $l_p = g^{1/3} l_s$ and the radius
of the M-circle $R_{s} = gl_s$,  with $g$ being the IIA string coupling 
constant. In the weak coupling region $g \ll 1$ of the IIA string, 
the three scales line up as
\beq
      R_{s} \ll l_p \ll l_s.
\eeq
In the low-energy short-distance region,
%\beq
%    E \sim \epsilon_{DKPS}= \frac{R_{11}}{l_p^2} \ll \frac{1}{l_p}, ~~
%    L \ll l_s,
%\label{dkpsregion}
%\eeq
the dynamics of D(irichlet) branes \cite{pol} is described by 
a gauge theory on the D-brane worldvolume \cite{d0brane,dkps,sen,sei}. In 
flat space, the $v^4$-term in the scattering amplitude of two
D$0$ branes computed in this short-distance region 
coincides with that computed in the supergravity region 
\cite{dkps,bbbb,bfss}.
%\beq
%    E \ll \frac{1}{l_p},~~ l_p \ll L, R_{11}.
%\label{sugraregion} 
%\eeq
In this case the corresponding gauge theory has sixteen supercharges,
and they protect the scattering amplitude from corrections that might 
arise when extrapolating the length scale.                                  
                   Generally speaking, amplitudes that are not protected 
by a sufficient number of supersymmetries
 may behave differently in the two
regions. Indeed in \cite{dos,do} it is shown that the $v^4$-term in the 
D$0$-brane scattering amplitude in a curved space (more specifically 
the large volume limit of K$3$, called the ALE space)
computed in the gauge theory region differs from
the amplitude computed in the supergravity region. The former
lacks subleading terms in the weak curvature expansion of 
the latter. Issues closely related to this were pointed out  
in \cite{dine,hp}. 

It is interesting to study other amplitudes in the gauge theory region and
compare them with those in evaluated in the supergravity region. 
In \cite{pp}, Polchinski and Pouliot computed membrane scattering in
flat space which an exchange of a  non-zero amount of M-momentum $p_{11}$.
The non-zero $p_{11}$ transfer corresponds to an instanton process
in the three-dimensional $N=8$ gauge theory (with sixteen supercharges)
on the membrane worldvolume. They showed that the leading term in the 
one-instanton calculation agrees with the leading term in 
the supergravity side for one unit of M-momentum exchange in  
the region $b \gg R_{s} \gamma$, where $b$ is the impact
parameter and $\gamma = (1-v_{11}^2)^{-1/2}$ is the Lorentz
boost factor for longitudinal velocity $v_{11}$. In view of the
result in \cite{do}, one may suspect that such an agreement does not
persist in a curved space. 

In this paper, we study membrane scattering on an ALE
space with non-zero $p_{11}$ transfer in the gauge theory region using the
$N=4$ gauge theory (with eight superchages) in three
dimensions.  In curved space, there is an 
issue of which gauge theory should be used to describe the
dynamics in the short-distance region \cite{douglas,dgstring,dko}.
In the case of membranes, the renormalizability 
of the corresponding three-dimensional gauge theory
imposes a strong constraint. As far as we know, the only reasonable
candidate is the gauge theory studied in \cite{dm,tensor,intsei,DHOO} 
whose field content and gauge groups are arranged according to 
the quiver diagram. In this paper, we call this the quiver gauge
theory. In the orbifold limit of the ALE space, the model can be
derived from open string dynamics \cite{dm}, and one can argue that
turning on Fayet-Iliopoulos parameters in field theory corresponds
to resolving the ALE space.

The instanton in three dimensions is a magnetic monopole. 
In the case studied by Polchinski and Pouliot \cite{pp}, the gauge
theory has $N=8$ supersymmetry, half of which is preserved by
the monopole. Consequently, there are eight fermion zero modes
in the instanton background, which is exactly
the right number to compute the $v^4$-term in the membrane scattering
amplitude, as the superpartner of the $F^4$ term in the gauge theory contains
eight fermions. In our case, we find that the monopole in question
breaks all the supersymmetry. Since our gauge theory has only $N=4$ 
supersymmetry, one can expect that the number of fermion zero-modes is
still eight. We will verify that this is in fact the case. 
This fact helps us in estimating the normalization of
the leading term in the instanton computation. 

However the fact that the instanton breaks all the supersymmetry 
makes it difficult to construct it explicitly and evaluate its
action. We address this problem in the following way.  We first construct 
a field configuration obeying the $n$-monopole boundary condition, 
and show that the action for this trial field configuration is less
than $n/e^2$ times the geodesic distance $\sigma(X,Y)$ between the membranes
($e$: gauge coupling constant). 
Since the instanton minimizes the action for the given boundary condition,
the
 instanton action $S_{instanton}$ should clearly
obey the same upper-bound. We also show that, for any field configuration
with the $n$-monopole boundary condition, the action is bounded below by 
$n/e^2$ times
the Euclidean distance $\parallel X-Y \parallel$ 
in the total field space. Thus we obtain the inequalities,
\beq
\label{aaa}
    \frac{n}{e^2}\parallel X-Y\parallel < S_{instanton} <
 \frac{n}{e^2} \sigma(X,Y),
\eeq
except in the orbifold limit when the inequalities above are
replaced by equalities. 
The agreement with the supergravity
computation would require the action to be exactly equal to 
$\frac{n}{e^2} \sigma(X,Y)$. Thus the gauge theory amplitude is 
exponentially larger the corresponding supergravity amplitude. 
Moreover we show that the error 
$\frac{n}{e^2} \sigma(X,Y) - S_{instanton}$ grows at least
linearly in $n$ for large $n$.

The membrane scattering amplitude computed using the gauge 
theory is different from the supergravity result. 
What does this imply to the  
large-$N$ Matrix Theory conjecture of Banks, Fischler,
Shenker and Susskind \cite{bfss}? 
To understand the origin of the discrepancy, we examine the
region of validity of the gauge theory computation within the framework 
of the large-$N$ conjecture.
In Matrix Theory, the membrane arises from 
a collection of a large number of D$0$-branes \cite{dhn}. The question is,
in this set-up, when it is appropriate to use 
the three-dimensional gauge theory to describe 
the membrane dynamics. If the number $N$ of
D$0$-branes is finite, there is a short-distance cutoff 
$\delta x \sim 1/\sqrt{N}$
on the worldvolume of the membrane. In order for the gauge theory 
description to be valid, the length scale of the problem has to be larger
than the cutoff length. We find that this requires
that the Planck length $l_p$ and the distance $\sigma(X,Y)$ between the
membranes be {\it shorter} than $\delta x$, 
\beq
     l_p, \sigma(X,Y) \ll \delta x.
\eeq   
Therefore, even in the large-$N$ Matrix
Theory, the gauge theory description of the membrane is valid only
in the short-distance regime.
The gauge theory and supergravity descriptions
cover complementary regimes and 
there is no overlap between the two. 
The result of this paper is an explicit
example where the gauge theory computation
is not applicable to the long-distance physics of membrane. 

In section 2, we define the gauge theory on the membrane worldvolume
and show that the instanton breaks all supersymmetries. In section
3, we prove the inequalities (\ref{aaa}) and estimate the error. 
Section 4 is devoted to an analysis of the zero-modes
in the instanton background. In section
5, we compare the gauge theory result with the supergravity prediction. 
The last section is devoted to a discussion on the large-$N$
Matrix Theory conjecture. In the appendix, we review
how the three-dimensional gauge theory description 
of membranes is derived starting from a system of $N$ D0 branes, and 
estimate the cutoff length of the gauge theory. 

\section{Quiver Gauge Theory for Membrane on ALE Space}

In Matrix Theory \cite{bfss}, the membrane arises from 
a collection of a large number of D$0$-branes \cite{dhn}.
The dynamics of the membrane is described by a
supersymmetric gauge theory on the worldvolume \cite{bss,vvk}. 
In this section, we define the gauge theory for two membranes on 
the ALE space. The derivation of the action is reviewed in the appendix.
We will examine the validity of this description in section~6 
of this
paper. 

According to \cite{pp,vvk}, the coupling constant $e$ 
of the gauge theory is given by the string computing constant $g_s$
as $e^2 = \gamma g_s/l_s$. The Lorentz boost factor
$\gamma$ is given by
\beq
  \gamma = \frac{N/R_s}{L^2/l_p^3}, 
\eeq
where $N$ is the number of D$0$-branes and $L$ is the 
size of the membrane.
In three dimensions, the gauge theory is weakly coupled if $e^2$ 
is smaller than the vacuum expectation value $|\phi|$ of the Higgs field. 
In this regime, the loop corrections are suppressed by powers of
$e^2/|\phi|$, and the instanton approximation is useful. 
In our case, $\phi$ is typically related to the distance between the
membranes $\sigma(X,Y)$ as
\beq
  |\phi| = \frac{\sigma(X,Y)}{l_s^2}.
\eeq
Therefore the gauge theory is weakly coupled when
\beq
  \frac{e^2}{|\phi|} = \frac{R_s \gamma}{\sigma(X,Y)} \ll 1.
\label{weakregime}
\eeq
In section 6, we will show that, 
if the gauge theory description of the membrane is valid, there is such
a weakly coupled regime in the parameter space. 

To simplify our notation, in this and next sections, we will set 
the string scale $l_s=1$.

\subsection{Construction of the Quiver Gauge Theory}

\newcommand{\vz}{\vec{\zeta}}

Let us start with a single flat membrane localized at a point in
$S^1 \times R^3 \times {\cal M}$, where 
${\cal M}$ is an ALE space of type $A_{n-1}$.
According to \cite{dm}, the dynamics of D0 branes propagating on
an ALE space is given by a one-dimensional quiver quantum mechanics.
Using this and the derivation of the membrane action from the D0
brane action in the appendix,
we find that the worldvolume dynamics
of the membrane is described by a three-dimensional $N=4$
supersymmetric gauge theory with gauge group
$U(1)^n=\prod_{i=0}^{n-1}U(1)_i$ and one massless hypermultiplet
charged $(+1,-1)$ with respect to
 $U(1)_i\times U(1)_{i+1}$ for each $i=0,...,n-1$
(where $U(1)_n\equiv U(1)_0$).
The Fayet-Iliopoulos (FI) parameter
$\vz=(\vz_0,\ldots,\vz_{n-1})$ is chosen so that $\sum_i\vz_i=0$.
The Higgs branch is the ALE space ${\cal M}$ which depends on
the FI parameter $\vz$. When $\vz=0$, the ALE space has an $A_{n-1}$
singularity, which is resolved for $\vz \neq 0$. 
For generic values of $\vz$, the moduli space consists of 
a single mixed Higgs-Coulomb branch $S^1 \times R^3 \times {\cal
M}$, which is identified as the space transverse to the membrane.
 
To describe two parallel membranes, we consider the gauge theory
with gauge group $U(2)^n = \prod_{i=0}^{n-1} U(2)_i$
and one hypermultiplet in $(2_i, \bar{2}_{i+1})$ 
of $U(2)_i \times U(2)_{i+1}$
for each $i=0,...,n-1$. 
%The gauge coupling constants for all the $U(2)_i$'s are chosen 
%to be equal as
%$e^2=2\gamma g/l_s$ where $\gamma = (1 -v_{11}^2)^{-1/2}$
%is the boost parameter \cite{pp}.
The moduli space of vacua of this theory was analyzed in \cite{DHOO}
and found to have various branches. 
The branch relevant in this paper 
is the mixed branch (called the basic branch in
\cite{DHOO}, see equation (5.51))
in which the gauge group is broken to the $U(1)\times U(1)$ subgroup
of the diagonal $U(2)\subset U(2)^n$.
This branch receives no quantum corrections and
is given by the two-fold symmetric
product of $S^1 \times R^3 \times {\cal M}$,
which precisely matches the picture of two membranes at points on
the transverse part of the space-time. The gauge coupling constant
is equal to $e \sqrt{n}$ for all $U(2)_i$ so that the coupling constant
for the diagonal $U(2)$ is $e$.

\subsection{Monopole and Supersymmetry}

To describe membranes moving with a large velocity in the eleventh
direction we need to turn on a magnetic flux in the three-dimensional
gauge theory. The momentum density $\pi_{11}$
of each membrane in the eleventh
direction is given by the magnetic flux $F_{12}/2\pi R_{s}$
of the corresponding unbroken $U(1)$ subgroup.
Therefore,
in a scattering with one unit of momentum transfer,
the magnetic flux $\int \dd x^1\dd x^2 F_{12}$
of one $U(1)$ increases by $2\pi$ and the flux of the other $U(1)$
decreases by the same amount.
This is realized as the magnetic monopole of the $SU(2)$ subgroup of
the diagonal $U(2)$. Namely, this is an instanton process of this
$SU(2)$.

Later we will need to know the
number of fermion zero modes in this instanton background. 
For this, it is useful to count
the number of unbroken supersymmetries.
In fact, the instanton in question breaks all the
supersymmetry. The easiest way to see this is to realize
the gauge theory using intersecting branes \cite{HW}. 
The brane configuration consists of 
D$3$ and NS$5$-branes in type IIB theory on the
flat space-time $R^9\times S^1$ which we parametrize by $x^0,\ldots,
x^9$ where the $S^1$ is in the $x^6$ direction \cite{DHOOY}.
There are $n$ parallel NS$5$-branes spanning
the $x^{0,1,2,3,4,5}$ directions and two parallel D$3$-branes spanning
the $x^{0,1,2,6}$ directions. Since the $x^6$ direction
is finite the worldvolume dynamics of the D$3$-branes
is at low energies described by a three-dimensional gauge theory. 
An analysis of the open string states
shows that the theory has the same gauge symmetry and
matter content as the quiver gauge theory given above.

Before we count the number of unbroken supersymmetries in the
instanton background, it is useful to briefly review  why this
configuration has $N=4$ supersymmetry in three dimensions. 
Type IIB theory in flat space-time has
$32$ supercharges parametrized by a pair of Majorana-Weyl spinors, 
$\epsilon_+$ and $\epsilon_-$, of the same chirality
$\Gamma_{0\ldots 9}\epsilon_{\pm}=\epsilon_{\pm}$.
The NS$5$-branes preserve half of the supersymmetries, namely the
ones satisfying
$\Gamma_{012345}\epsilon_{\pm}=\pm\epsilon_{\pm}$,
and the D$3$ branes preserves those
that obey $\Gamma_{0126}\epsilon_+=\epsilon_-$. 
Therefore $32/(2\cdot 2)=8$ generators are preserved, which
is the number of supersymmetries of $N=4$ in three dimensions.

\begin{figure}[htb]
\begin{center}
\epsfxsize=3.5in\leavevmode\epsfbox{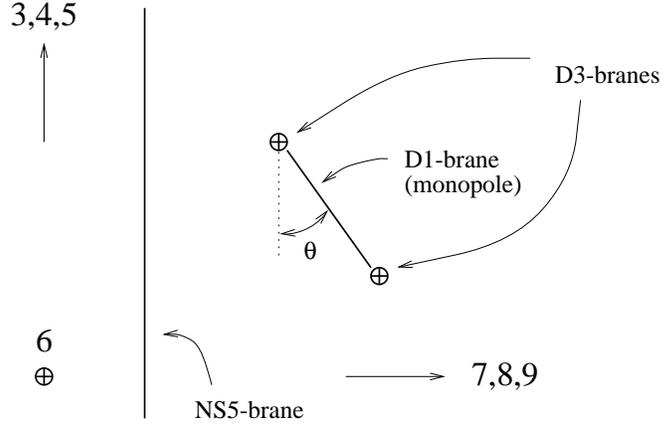}
\end{center}
\caption{The monopole in the Higgs or mixed Higgs-Coulomb branch 
($0< \theta < \pi$)  breaks all the
supersymmetry.}
\label{fig0}
\end{figure}

Now let us show that the instanton in question break all the
remaining supersymmetry. 
As noted in \cite{HW,DHOOY}, an instanton in the three-dimensional
gauge theory
is realized as the open Euclidean D$1$-brane 
stretched between the two D$3$-branes and in the $x^6$ direction.
If we were in the Coulomb branch, the two D$3$-branes would have been 
separated in the $x^{3,4,5}$ directions. Then the instanton 
realized as the D$1$-brane would, for D$3$-branes separated in $x^3$ only, 
impose the additional condition
$\Gamma_{36}\epsilon_{+}=\epsilon_-$ for  
unbroken supersymmetry.
Since $\Gamma_{36}$ anti-commutes with
$\Gamma_{012345}$, this would have been compatible with
the condition $\Gamma_{012345}\epsilon_{\pm}=\pm\epsilon_{\pm}$
associated with the NS$5$-branes, thus there would have been 
four unbroken supercharges. We are, however 
in the mixed Coulomb-Higgs branch. There  the two D$3$-branes are
separated in a mixed direction of $x^{3,4,5}$ and $x^{7,8,9}$,
say $\cos \theta \partial_3+\sin \theta \partial_7$ (see figure 1).
Thus the condition is $(\cos\theta\Gamma_{36}
+\sin\theta\Gamma_{76})\epsilon_+=\epsilon_-$. 
In this case, since $\Gamma_{76}$ commutes with $\Gamma_{012345}$,
it is not compatible with the condition from the NS 5-branes.
This proves that the supersymmetry is completely broken by
instantons in the Higgs ($\theta = \pi/2$) or mixed branch
($\theta \neq 0, \pi/2$).
If there were no NS$5$-branes,
the two conditions would not have been in conflict
and some of the supersymmetries would have been preserved.
This corresponds to the case of $N=8$ supersymmetry discussed
in \cite{pp}.

This has the following consequence in the corresponding
$N=2$ theories in four dimensions. Note that instantons in
three dimensions are solitons in four dimensions.
In four dimensions, the above observation translates 
into the statement that
monopoles in the Higgs or mixed branches, if they exist,
cannot be Bogomolny-Prasad-Sommerfeld (BPS) particles. 
As is well-known BPS states
are in the short multiplet of $N=2$ supersymmetry consisting
of four states, whereas non-BPS states are in the long multiplet
consisting of sixteen states. If the monopole
is not a BPS state, there must be a sufficient number of degrees of freedom
to make a long multiplet. In the Higgs (or mixed) branch,
one $N=2$ vector multiplets (eight states) mixes via the Higgs mechanism
with one hypermultiplet (eight states) and in total there are the
right number of states to make a long multiplet.
In the Coulomb branch, one vector multiplet mixes with itself
and decomposes into one BPS and one anti-BPS multiplet.
In the mixed branch of $N=4$ super-Yang-Mills theory,
the sixteen states mentioned above
decompose again into four BPS/anti-BPS multiplets.

\section{Monopole Solution and Action}

We have found that the monopole solution in our case
preserves no supersymmetry. This makes it difficult to find an exact form of
the solution. We address this issue by using the following strategy. 

First we start with the quiver gauge theory defined in the last section,
constrain almost all massive excitations of the scalar fields to be zero 
and go to the low-energy effective theory. The scalar fields are 
subject to the D and F-term constraints and are only allowed to move 
along the moduli space of vacua. In this model, we construct a
trial      field 
configuration obeying the boundary condition for the $n$-monopole, and
show that its action is smaller than $n/e^2$ times
the geodesic distance
$\sigma(X,Y)$ between the two membranes. Since the instanton minimizes
the action, the instanton action should also obey the same
upper-bound. We also show that, for any field configuration obeying this
boundary condition, the action
 is bounded below by $n/e^2$
 times the Euclidean
distance $\parallel X - Y \parallel$ 
between the two membranes. Thus the instanton 
action for the $n$-instanton should be bounded by these two quantities
as
\beq
      \frac{n}{e^2}\parallel X-Y\parallel < S_{instanton} < 
 \frac{n}{e^2} \sigma(X,Y),
\label{bound}
\eeq
except in the orbifold limit of the ALE space ${\cal M}$ when the
inequalities are replaced by equalities.  In the following, we assume
${\cal M}$ is smooth (i.e. $\vz \neq \vec{0}$) unless otherwise
noted. 

We then proceed by showing
 that the above field configuration can be lifted to the
full quiver gauge theory. To solve
the equations of motion of the full theory, we also have to minimize
the action with respect to the massive modes. 
We show that this further increases the amount of the error, 
$\frac{n}{e^2}
\sigma(X,Y) - S_{instanton}$, although the action still obeys  the 
lower-bound in (\ref{bound}).

\bigskip
\noindent
\subsection{Monopole Configuration in the Low Energy Theory}

The low energy bosonic degrees of freedoms of the 
gauge theory are the scalar fields $(X^i,Y^i)$ $(i=1,\ldots,7)$
taking values in the
moduli space of vacua ${\rm Sym}^2(R^3 \times {\cal M})$ and
the gauge field $a_\mu^a$ ($a=1,2$) of the $U(1)\times U(1)$
subgroup of the diagonal $U(2)$ and are described by the
Lagrangian density,
\beq
e^2{\cal L}_{eff}={1\over 4}\sum_{a=1,2}(\partial_{[\mu,} a_{\nu]}^a)^2
+{1\over 2}g_{ij}(X) \partial_\mu X^i \partial_\mu X^j + {1\over 2}
g_{ij}(Y) \partial_\mu Y^i \partial_\mu Y^j,
\eeq
where $g_{ij}$ is the metric of the space $R^3\times{\cal M}$.

In order to construct the monopole, we need to
retain the W-bosons $A_{\mu}^{\pm}$
of the diagonal $U(2)$ which are generically massive
on the moduli space ${\rm Sym}^2(R^3 \times {\cal M})$ 
but become massless at the diagonal of the symmetric product where
the two membranes coincide.
We therefore consider the
$U(2)$ gauge theory\footnote{The action in (\ref{effect}) is
actually the $U(2)$ gauge theory in a specific unitary gauge,
namely one where all scalar fields are diagonal. (In the quiver
gauge theory, all scalar fields transform in the adjoint of the
diagonal $U(2)$). By introducing an additional `compensating'
 $U(2)$ valued field one can make (\ref{effect}) into a gauge theory
where the gauge has not yet been fixed. The unitary gauge chosen
here is equivalent to choosing the gauge $\phi={\rm diagonal}$ in
the usual BPS equation $F=\ast D \phi$. For the BPS monopole solution,
this gauge cannot be chosen globally, but only on the two
hemispheres of a sphere at infinity. Along the equator, one has
to glue the solutions together using a non-trivial transition function.
In principle we would have to do the same thing for the trial monopole
configuration that we will construct, but it is easy to see that exactly
the same transition functions can be used to promote our trial 
configuration into an everywhere well-defined configuration.}
 with the following Lagrangian density,
\beq
e^2{\cal L}_{eff} ={1\over 4} {\rm tr}~ F_{\mu\nu}^2
+ {1\over 2}g_{ij}(X) 
\partial_\mu X^i \partial_\mu X^j +
{1\over 2}g_{ij}(Y) 
\partial_\mu Y^i \partial_\mu Y^j + 2\parallel X-Y\parallel^2
A_\mu^+ A_\mu^-
\label{effect}
\eeq
where $\parallel X-Y\parallel$ is the W-boson mass given as follows. 
Recall that the original quiver gauge theory has three real
scalar fields $\vec{\varphi}_i$ in the adjoint representation
for each $U(2)_i$,
and a pair of complex scalars 
$(Q_i, \tilde{Q}_i)$ 
which is in $(2_i, \bar{2}_{i+1})$ for each
$U(2)_i \times U(2)_{i+1}$.  
At each point of the moduli space $(X,Y)$,
these matrix-valued scalars can be diagonalized 
\cite{DHOO} as
\beqa
\vec{\varphi}_i= \left( \begin{array}{cc}
 \vec{\varphi}(X) & 0 \\ 
 0 & \vec{\varphi}(Y) \end{array} \right),~~
Q_i= \left( \begin{array}{cc} 
  b_i(X) & 0 \\ 0 & b_i(Y) \end{array} \right),~~
\tilQ_i= \left( \begin{array}{cc}
   \tilde{b}_i(X) & 0 \\ 0 & \tilde{b}_i(Y) 
\end{array} \right).
\label{reps}
\eeqa
Here both $(b_i(X),\tilde{b}_i(X))$ and
$(b_i(Y),\tilde{b}_i(Y))$ obey the D and F-term constraints
of the $\prod_i U(1)_i$ gauge theory
for a single membrane located at $X$ and $Y$ respectively. 
The Euclidean distance $\parallel X-Y\parallel^2$ is then
defined as
\beq
\parallel X-Y\parallel^2
\,\,\equiv\,\,\,n |\vec{\varphi}(X)-\vec{\varphi}(Y)|^2
+\sum_{i}(|b_i(X)-b_i(Y)|^2+
|\tilde{b}_i(X)-\tilde{b}_i(Y)|^2).
\label{diag}
\eeq
In other words, $\parallel X-Y\parallel$ is the Euclidean distance
of some representative points of $X$ and $Y$ in the vector space
for the scalar fields of the $\prod_i U(1)_i$ 
gauge theory for a single
membrane. That the W-boson mass of the low energy theory is equal to
the Euclidean distance in the total field space of the quiver gauge
theory was pointed out in \cite{dos}.  
The value of the Euclidean distance clearly depends on how to take the
representative points (\ref{reps}). Below we will
specify how it is done in our case. 

Let us first rewrite the action (\ref{effect}). 
For a function $f(X,Y)$ on ${\rm Sym}^2(R^3 \times {\cal M})$, define
\beq
  \Lambda = \left( \begin{array}{cc} 
                 f(X,Y) & 0 \\
                 0 & -f(X,Y) \end{array} \right). 
\eeq
We can use this to recombine the kinetic term for the gauge field as
\beq
  {\rm tr}~ F_{\mu\nu}^2 =
  {\rm tr}~ (F_{\mu\nu} - \epsilon_{\mu\nu\rho} D_\rho \Lambda)^2
 + 2 d~ {\rm tr}~ F \Lambda - 2(\partial_\mu f)^2 - 8 f(X,Y)^2
 A_\mu^+ A_\mu^-. 
\eeq
The effective Lagrangian density (\ref{effect}) then  becomes
\beqa
 e^2{\cal L}_{eff} &=& {1\over 4}{\rm tr}~(F_{\mu\nu} 
 - \epsilon_{\mu\nu\rho} D_\rho \Lambda )^2 
 + {1\over 2}(\partial_\mu X^i, \partial_\mu Y^i) M_{ij}
 \left( \begin{array}{c} \partial_\mu X^j \\ \partial_\mu Y^j 
 \end{array} \right) + \nonumber \\
&~& + 
  {1\over 2}d ~{\rm tr}~ F \Lambda + 2\Big[
\parallel X-Y\parallel^2 - f(X,Y)^2 \Big]
 A_\mu^+ A_\mu^-. 
\label{monopoletrick}
\eeqa
where the matrix $M_{ij}$ is given by
\beq
 M_{ij} = \left( \begin{array}{cc} 
     g_{ij}(X) - \frac{1}{2} \frac{\partial f}{\partial X^i}
            \frac{\partial f}{\partial X^j} &
 -\frac{1}{2}\frac{\partial f}{\partial X^i}
                            \frac{\partial f}{\partial Y^j} \\
-\frac{1}{2}\frac{\partial f}{\partial Y^i}
                            \frac{\partial f}{\partial X^j}
& g_{ij}(Y) - \frac{1}{2}\frac{\partial f}{\partial Y^i}
                            \frac{\partial f}{\partial Y^j}
\end{array}
\right).
\eeq

We now construct a field configuration which sets the first two terms
in the Lagrangian density (\ref{monopoletrick}) to be zero. This can be done by
making use of the standard BPS monopole solution to 
the $SU(2)$ gauge theory with a single adjoint scalar  
field $\Phi(x)$. The BPS equation for them is
\beq
  F_{\mu\nu} = \epsilon_{\mu\nu\rho} D_\rho \Phi. 
\label{bps}
\eeq 
A general $n$-monopole solution to (\ref{bps}) is given
by Corrigan and Goddard \cite{cg}. By choosing our $U(2)$
gauge field $A_\mu$ to be equal the $n$-monopole 
solution to (\ref{bps}) and by requiring 
the massless scalars $(X,Y)$ to obey
\beq
  f(X(x),Y(x)) = \phi(x),~~~(\pm \phi {\rm :~eigenvalues~of~} \Phi),
\label{phieq}
\eeq
we can set the first term in the Lagrangian density
 (\ref{monopoletrick}) to be
zero. 

So far the function $f(X,Y)$ has been arbitrary. 
To obtain the upper bound in (\ref{bound}) 
we choose the function to be the geodesic distance between $X$ and $Y$,
\beq
   f(X,Y)= \sigma(X,Y).
\eeq
With this choice, the matrix $M_{ij}$ is positive semi-definite. This
follows from the Hamilton-Jacobi equation for the geodesic distance
\beq
    g^{ij}(X) \frac{\partial \sigma}{\partial X^i}
                            \frac{\partial \sigma}{\partial X^j}
     + g^{ij}(Y) \frac{\partial \sigma}{\partial Y^i}
                            \frac{\partial \sigma}{\partial Y^j} = 1.
\label{hj}
\eeq
The second term in the Lagrangian density
(\ref{monopoletrick}) can be set to be zero if $(\partial_\mu X^i,
\partial_\mu Y^i)$ is the zero eigenvector of the matrix $M_{ij}$, namely
\beqa
    g_{ij}(X) \partial_\mu X^j &=& \frac{\partial \sigma}{\partial X^i}
  \partial_\mu \sigma(X(x),Y(x)) \nonumber \\
    g_{ij}(Y) \partial_\mu Y^j &=&  \frac{\partial \sigma}{\partial Y^i}
  \partial_\mu \sigma(X(x),Y(x)). 
\label{geodesics}
\eeqa
To solve this equation, we use the following property of geodesics.
Suppose $(X_0(\tau), Y_0(\tau))$ is a
geodesics on ${\rm Sym}^2(R^3 \times {\cal M})$ with $\tau$ being the
proper time and with the initial condition
\beq
   X_0 = Y_0 ~~~{\rm at}~\tau=0. 
\eeq
Because of the property of geodesics,
\beq
   g_{ij}(X_0) \frac{dX_0^j}{d\tau} 
=  \frac{\partial \sigma(X_0,Y_0)}{\partial
X_0^i},~~
 g_{ij}(Y_0) \frac{dY_0^j}{d\tau} =  
\frac{\partial \sigma(X_0,Y_0)}{\partial Y_0^i}.
\eeq
the equations (\ref{phieq}) and  (\ref{geodesics}) are 
both satisfied by setting
\beq
  X(x) = X_0(\phi(x)),~~Y(x) = Y_0(\phi(x)). 
\label{xy}
\eeq

Thus we find that, for the field configuration 
given by (\ref{bps}) and (\ref{xy}),
the Lagrangian density (\ref{monopoletrick}) becomes
\beq
  e^2 {\cal L}_{eff} =  
{1\over 2} d~ {\rm tr}~ F \Lambda + 2\Big[ \parallel X-Y\parallel^2
 - \sigma(X,Y)^2 \Big]
 A_\mu^+ A_\mu^-.
\label{eqqq}
\eeq
The second term in the right-hand side is negative definite
for the following reason. As we will show in the next 
subsection, in order to lift the above
monopole solution to the original quiver gauge theory,
we must choose a gauge slice for the $\prod_i U(2)_i$ gauge symmetry
that is normal to the gauge variation at any value of $\tau$.
For such a choice, the hyper-K\"ahler quotient
metric induced on the geodesic is the same as the restriction of the
flat Euclidean metric to the gauge slice,
and the geodesic length $\sigma(X,Y)$ is the same as the length
of the slice measured by the flat metric.
Since $\parallel X-Y\parallel$ is the length of the straight
segment connecting the same end points measured by the
same flat metric, this must be smaller than the length
of the slice which is equal to $\sigma(X,Y)$. 
Since the first term of (\ref{eqqq})
is a total derivative, its integral is determined
by the $n$-monopole boundary condition. Thus we
 find that the action for the field configuration
(\ref{bps}), (\ref{xy})
is bounded above by the geodesic distance as
\beq
   S_{eff}=\int d^3 x {\cal L}_{eff} 
 < \frac{n}{e^2}  \sigma(X,Y)_{|x \rightarrow\infty}
\eeq
Since the instanton minimizes the action, $S_{instanton}$ is smaller
than the action $S_{eff}$ for the trial monopole configuration
considered here. Thus it should also obey the same upper bound. 

We can also use (\ref{monopoletrick}) to find a lower bound on the
instanton action. For this, we set $f(X,Y)$ to be the Euclidean distance 
$\parallel X-Y\parallel$. The Lagrangian density then becomes
\beq
  e^2 {\cal L}_{eff} = 
{1\over 4} {\rm tr}~ (F_{\mu\nu} 
 - \epsilon_{\mu\nu\rho} D_\rho \Lambda)^2 
 + {1\over 2}(\partial_\mu X^i, \partial_\mu Y^i) M_{ij}
 \left( \begin{array}{c} \partial_\mu X^j \\ \partial_\mu Y^j 
 \end{array} \right) + 
  {1\over 2} d~ {\rm tr}~ F \Lambda .
\eeq
It is easy to see that the matrix $M_{ij}$ is now strictly positive
definite and the second term in the right-hand side is positive. 
We thus obtain the lower bound
\beq
\frac{n}{e^2} \parallel X-Y \parallel <  S_{eff}
\eeq
for any field configuration satisfying the $n$-monopole boundary
condition. 

\subsection{Estimate of the Error}

Let us estimate the size of the error, 
\beq
   \frac{n}{e^2} \sigma(X,Y) - S_{eff}
  = \frac{2}{e^2}
 \int d^3 x  \Big[ \sigma(X(x),Y(x))^2- \parallel X(x) - Y(x) 
\parallel^2 \Big] A_\mu^+ A_\mu^-.
\label{error}
\eeq
When $n=1$, the main contribution
to the error comes from the region where the distance $r$ from the
center of monopole is less than the Compton wavelength of 
the W-boson. When the distance between $X$ and $Y$ is shorter than
the typical curvature radius $R_c$ of ${\cal M}$, the difference
between the geodesic distance and the Euclidean distance is estimated as
\beq
   \sigma(X,Y)^2 - \parallel X - Y \parallel^2 \sim
    \frac{\sigma^4(X,Y)}{R_c^2}.
\eeq
In our monopole configuration,
\beqa
  \sigma(X(r),Y(r)) &=& \phi(r) \nonumber \\
  &=& \sigma {\rm coth}(\sigma r) - \frac{1}{r},
\eeqa  
where $\sigma$ in the right-hand side
is the value of $\sigma(X,Y)$ at $r=\infty$. By
combining with the known expression for the W-bosons $A^\pm$  in the BPS
solution, we find
\beq
 \Big[ \sigma^2(X,Y)- \parallel X - Y
\parallel^2 \Big] A_\mu^+ A_\mu^- \sim 
 \frac{\sigma^6}{R_c^2} F(\sigma r),
\eeq
where $F(\xi)$ is a certain function which decays exponentially for
large $\xi$. The error (\ref{error}) for $n=1$ is then estimated as
\beq
  \frac{1}{e^2} \sigma(X,Y) - S_{eff}
 \sim \frac{\sigma^3}{e^2 R_c^2}
 \int_0^\infty d\xi \xi^2 F(\xi).
\label{singlemono}
\eeq
This in fact is of the same order as the difference of the geodesic
distance and the Euclidean distance between $X$ and $Y$. Because of
the exponential suppression by $F(\sigma r)$, 
the main contribution to the integral (\ref{error}) is from the region
within the Compton wavelength of the W-boson. 
Since the instanton action $S_{instanton}$ is in general smaller than
the trial monopole configuration, the actual error
$\frac{1}{e^2} \sigma - S_{instanton}$ for the instanton action 
can be larger than the one given by (\ref{singlemono}). 

Now let us estimate the magnitude of the error for
an arbitrary value of $n$. For this, it is easiest to
consider the case of the well-separated BPS monopoles as our trial
configuration. If the distance
between monopoles is larger than the Compton wavelength of the
W-boson, the behavior of the $n$-monopole solution near the core of
each monopole
can be approximated by that of the single monopole solution. Since the
integrand of (\ref{error}) is positive definite, there is no
possibility of cancellation of the errors. 
Thus for the well-separated $n$-monopole,
\beq
  \frac{n}{e^2} \sigma(X,Y) - S_{eff}
 \sim \frac{n\sigma^3}{e^2 R_c^2}
 \int_0^1 d\xi \xi^2 F(\xi).
\eeq
Since the instanton action is smaller than that of the trial
configuration, the actual error can be larger than this. Thus we find
that the difference between the geodesic distance
$\frac{n}{e^2} \sigma(X,Y)$ and the instanton action $S_{instanton}$
is at least of the order $n$ for large $n$.

\bigskip

\noindent 

\subsection{Lifting to the Quiver Gauge Theory}

The low-energy theory with the Lagrangian density (\ref{effect}) is not
renormalizable in three dimensions. It could still be used for
computations if the energy scale of the problem is
smaller than that set by the
curvature of the moduli space ${\cal M}$. The energy scale
for the instanton solution is determined by
how fast the fields $A_\mu$, $X$ and 
$Y$ change on the membrane worldvolume and it is controlled by 
the values of $X$ and $Y$ at the infinity, typically by the distance
$\sigma(X,Y)$ between the membranes. Thus we expect that computations
done in the low-energy theory (\ref{effect}) suffer large corrections
when the distance $\sigma(X,Y)$ is comparable to the curvature radius
of ${\cal M}$. In particular, for the purpose 
of computing subleading terms 
in the weak curvature expansion of the scattering amplitudes, 
the analysis using the low-energy theory is not sufficient and we have
to use the renormalizable gauge theory discussed in section 2. 

It turns out that the
instanton action for the full quiver gauge theory obeys the same bound
(\ref{bound}) as the instanton action for the low-energy effective
theory. In this subsection, we show that the field 
configuration constructed
in the previous subsection can be lifted to that of the quiver
gauge theory and that the value of the action for the lifted configuration
remains the same. Thus the upper-bound for the instanton action 
given by (\ref{bound}) also holds for the quiver gauge
theory. In the next subsection, we show that the instanton action for
the quiver gauge theory is in fact lower than that of the low energy
theory, but is still bounded from below as in (\ref{bound}). 

First let us briefly review the procedure  to obtain
the non-linear sigma model on the K\"ahler quotient
as the classical low-energy effective theory of the quiver gauge theory.
(The sigma model on the {\it hyper}-K\"ahler quotient can be obtained simply
by restricting to the zero of the F-term potential.)
Suppose we are given a complex scalar field $\Phi$ in a representation
of the gauge group $G$, with the Lagrangian density
\beq
{\cal L}={1\over 2e^2}\parallel F_{\mu\nu}\parallel^2
+\parallel D_{\mu}\Phi\parallel^2+
{e^2\over 2}\sum_{a=1}^{\dim G}|\Phi^{\dag}T_a\Phi-\zeta_a|^2.
\label{orig}
\eeq
The last term is the D-term potential where $T_a$'s are the
generators of the gauge group and $\zeta_a$ is the FI parameter
which has values in the center of the gauge group only.
We consider for simplicity only the region of the values of the
scalar field where the gauge group is completely broken.
The gauge field acquires a mass from the second term
and we  can first integrate out the gauge field.
The second term $\parallel D\Phi\parallel^2$ contains
terms linear and quadratic in the
gauge field and we can eliminate the
gauge field by completing the square, ignoring their kinetic terms
for the moment. 
The variation of this term with respect to the gauge field
is expressed, with the aid of the D-term constraint, as
$
\delta\parallel D\Phi\parallel^2
=-2\delta A_{\mu}^a\,\Phi^{\dag}T_aD_{\mu}\Phi.
$
It is extremized if
\beq
M_{\bar a b}A^b_{\mu}=-(T_a\Phi)^{\dag}\partial_{\mu}\Phi
\label{eqn}
\eeq
where $M_{\bar a b}=(T_a\Phi)^{\dag}T_b\Phi$
is a matrix which has an inverse $M^{b\bar a}$
if the gauge group is completely
broken by $\Phi$. Next, we solve (\ref{eqn})
with respect to $A_{\mu}^b$ and plug the
solution into $\parallel D\Phi\parallel^2$. We obtain
\beq
\left.\parallel D\Phi\parallel^2\right|_{\mbox{\tiny(\ref{eqn})}}
=
\partial_{\mu}\Phi^{\dag}\partial^\mu \Phi-
\Big[\partial_{\mu}\Phi^{\dag}T_b\Phi \Big]M^{b\bar
  a}\Big[(T_a\Phi)^{\dag}\partial_{\mu}\Phi\Big].
\eeq
This yields the non-linear sigma model Lagrangian density
$G_{i\bar \jmath}\partial_{\mu}
X^i\partial_{\mu}X^{\bar\jmath}$ on the K\"ahler quotient
for an arbitrary
choice of the gauge slice $X\to \Phi(X)$ in the D-constrained
manifold. 

Substituting (\ref{eqn}) into
the gauge kinetic term $\parallel F_{\mu\nu}\parallel^2$, we obtain
a higher derivative term for $\Phi$. Therefore
the resulting Lagrangian density
 would be in general different from the one given by (\ref{effect}). 
Fortunately this is not the case for 
the monopole solution constructed in the previous subsection. 
If we can choose a gauge slice $\Phi(X)$ so that
it is everywhere normal to
the gauge orbit, the right-hand side of (\ref{eqn}) vanishes. 
In general, such a slice may or may not exist. 
If it exists, the term in $\parallel D \Phi \parallel^2$
linear in $A_\mu$ vanishes for such a gauge slice and 
the second term of (\ref{orig}) becomes of the form
\beq
\parallel D\Phi\parallel^2
=
G_{i\bar\jmath}\partial_{\mu}X^i\partial_{\mu}X^{\bar\jmath}
+\parallel A_{\mu}\,\Phi(X)\parallel^2.
\label{kin}
\eeq
Therefore we can set the gauge field $A_\mu$ to be equal to zero 
and still get the non-linear sigma-model Lagrangian density.

In our quiver gauge theory, such a normal slice does not exist
for a generic configuration $(X(x),Y(y))$.
Fortunately our monopole solution
 is special in the sense that it evolves along a one-dimensional
subspace in the total field space. This is because
$(X,Y)$ depends on $x^{\mu}$ only through the single
function $\phi(x)$ as in (\ref{xy}). Therefore we can choose 
a normal slice for this
configuration and obtain (\ref{kin}) for the
kinetic term of the scalar field.
The normal section can be chosen so that 
all bifundamentals are simultaneously diagonalized as in (\ref{diag}).
If we keep only the gauge field of the diagonal $U(2)$ subgroup,
then the term quadratic in $A_{\mu}$ in (\ref{kin}) is
\beq
\parallel A_{\mu}\,\Phi(X,Y)\parallel^2
=\,\parallel\! X-Y\!\parallel^2 A_{\mu}^+A_{\mu}^-,
\eeq
where $\parallel X-Y\parallel^2$ is defined by (\ref{diag}).
Thus, the Lagrangian density 
of the quiver gauge theory for this configuration
is the same as
${\cal L}_{eff}$ in (\ref{effect}) if we set 
all the massive fields equal to zero, except for
the W-bosons in the diagonal $U(2)$ subgroup.

\bigskip

\noindent             
\subsection{Effects of Massive Modes}

We have shown that there is a particular configuration of the quiver
gauge theory, satisfying the boundary condition for the $n$-monopole,
such that its action is bounded from above by $\frac{n}{e^2}
\sigma(X,Y)$. The configuration, however, may not be a solution to the
full equations of motion of the quiver theory as we are simply setting
the massive fields to be equal to zero, except for the W-boson of the
diagonal $U(2)$. In this subsection, we will estimate the effects of
the massive modes. We will show that they increase the amount of the
error estimated in section 3.2.

When we turn on the massive modes of the scalar fields, the monopole
solution will deviate from the BPS configuration (\ref{bps}),
(\ref{xy}).  The massive modes are not negligible when the moduli
space embedded in the large field space has non-zero extrinsic
curvature. In such a case, the motion of the massless modes along the
moduli space in general generates a centrifugal force (or rather centripetal
force in this case as we are considering an Euclidean rather and
Minkowskian equation of motion), which will cause the trajectory to deviate
from the moduli space. The situation is illustrated in figure 2.
Here we will show that, unless the FI parameters $\vz$ are all equal
to zero, the centripetal force is non-zero. The case with $\vz=0$
corresponds to the orbifold limit of ${\cal M}$.
Thus, with the massive fields turned on, the monopole solution deviates
from the BPS configuration found in the previous section. This is an
additional difficulty in finding an exact form of the 
instanton solution. However, the very
fact that the actual solution deviates from the BPS solution means
that the action for the solution is smaller than the one evaluated in
the last two subsections.   

\begin{figure}[htb]
\begin{center}
\epsfxsize=1.5in\leavevmode\epsfbox{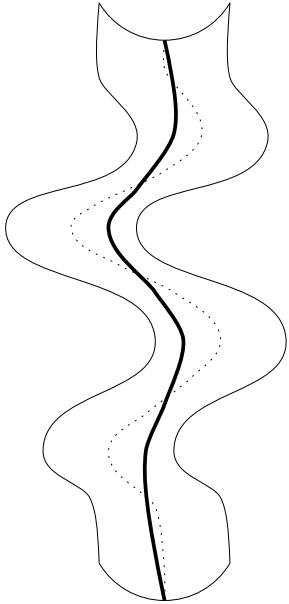}
\end{center}
\caption{The trajectory does not follow the ground state of the
potential due to the centripetal force.}
\label{fig1}
\end{figure}

Let us study how the solution deviates from the sigma-model 
solution.   
In the gauge theory described in section
2.1, there is one hypermultiplet field 
$(Q_i, \tilde{Q}_i)$ 
which is in $(2_i, \bar{2}_i)$ for each  
$U(2)_i \times U(2)_{i+1}$. The D and F-terms for 
each $U(2)_i$ are
\beq
  D^{A}_i = \sigma^{A}_{ab} 
     \left(  \Phi_{i-1}^{\dagger a} \Phi_{i-1}^{b}
       -  \bar{\Phi}_{i}^{a} \Phi_{i}^{\dagger b}
   \right) = \zeta_i^A {\bf 1}_i ,
\label{dterm}
\eeq
where $a,b=1,2$ are the hyperk\"ahler indices as
\beq
   (\Phi_{i}^{1}, \Phi_i^2)  = (Q_{i}, \tilde{Q}_i^*),
\eeq
and
$\sigma^{A=1,2,3}$ are the standard Pauli matrices. 
The potential for the scalar fields of the gauge theory is then
\beq
    V(\Phi) = \sum_{i} \sum_{A=1}^3
          {\rm tr} | D_{i}^{A} - \zeta_i^A {\bf 1}_i |^2, 
\label{potential}
\eeq
where ${\rm tr}$ is over the $U(2)_i$ gauge group representation. 

%If we require $V(\Phi)=0$, we obtain three constraints for each
%$U(2)_i$ and (\ref{dterm}) can be solved with 
%$ 4 \cdot 2 \cdot 2 \cdot N - 3 \cdot 4 \cdot N = 4N$ 
%parameters. Although there is an $U(2)^N$ gauge symmetry, the diagonal 
%$U(1) \times U(1)$ does not act on any of the hypermultiplet
%fields. Thus we are left with eight gauge invariant parameters, which
%correspond to the ${\cal M}$ part of $(X, Y)$ in section 2.2. 
%Thus at low energies, the gauge theory reduces to the 
%non-linear sigma-model on ${\rm Sym}^2 (R^3 \times {\cal M})$. 

Suppose the general solution to $V(\Phi)=0$ is parametrized by moduli
 fields  $\phi^\alpha$
as $\Phi_i^a = \Phi_{i(0)}^a(\phi)$.
 By construction, the BPS configuration
satisfies
\beq
 \left(
\frac{\partial D^{A}_{i}}{\partial \Phi_i^a} \right)_{|\Phi=\Phi_0}
\frac{\partial \Phi_{i(0)}^a}{\partial \phi^\alpha}  = 0,
\label{normal}
\eeq
namely normal vectors to
the moduli space 
are given by derivatives of the D-terms
$D_{i}^{A}$ with respect to $\Phi$, modulo gauge transformation. 
As the metric in the total
field space is flat, the centripetal force is given by
\beqa
F_{i}^{A} &=&  
 \left( \frac{\partial D^{A}_{i}}{\partial \Phi_i^a} 
    \right)_{|\Phi=\Phi_0}
\frac{\partial^2 \Phi_{i(0)}^a}{\partial \phi^\alpha 
\partial \phi^\beta} D_\mu \phi^\alpha D^\mu \phi^\beta 
 \nonumber \\
&=&  \sigma^{A}_{ab}
     \left(  D_\mu \Phi_{(0)i-1}^{\dagger a}
 D^\mu \Phi_{(0)i-1}^{b}
       -  D_\mu \bar{\Phi}_{(0)i}^{a}
 D^\mu \Phi_{(0)i}^{\dagger b}
   \right) . 
\label{centri}
\eeqa
Here we used (\ref{normal}) a couple of times to reduce the equation. 

Now we are ready to show that the centripetal force $F_i^A$ is in fact
non-zero for the solution constructed in the previous section, except
in the orbifold limit of ${\cal M}$. 
As the fields $\Phi$ in the monopole configuration depend on $x$
through the propertime $\tau = \phi(x)$, the vanishing of the force
$F_i^a=0$ would require
\beq
   \sigma^{A}_{ab}
     \left(  D_\tau \Phi_{(0)i-1}^{\dagger a}
 D_\tau \Phi_{(0)i-1}^{b}
       -  D_\tau \bar{\Phi}_{(0)i}^{a}
 D_\tau \Phi_{(0)i}^{\dagger b}
   \right) =0. 
\eeq
This, however, is the same as the D and F-term constraints for
${\rm Sym}^2( R^3 \times {\cal M})$ with $\vz=0$. 
This means that $D_r \Phi_{(0)}$ is on the orbifold. 
It is straightforward to show that this is possible only if $\Phi_{(0)}$
itself is on the orbifold. This means $\vz=0$. Thus we find that
the centripetal force is non-zero unless $\vz=0$.

On the other hand, the action is clearly larger than the Euclidean
distance in the total field space. To see this, suppose we remove the
D an F-term potentials from the Lagrangian density. We then have the
$\prod_i U(2)_i$ gauge theory coupled to massless scalar fields. By using the
standard BPS argument, one can easily show that the action for the
field configuration obeying the $n$-monopole boundary condition
 is bounded below by $\frac{n}{e^2} \parallel X - Y \parallel$. Adding
the D and F-term potentials back would simply increase the value of
the action. Thus the action of the quiver gauge theory is also bounded
below by the Euclidean distance. 

This concludes the proof of the inequality 
\beq
   \frac{n}{e^2} \parallel X - Y \parallel^2 <
S_{instanton} < \frac{n}{e^2} \sigma(X,Y),
\eeq
and the estimate of the error $\frac{n}{e^2} \sigma(X,Y)
-S_{instanton}$.

%This result is in sharp contract to the mass of the
%ground state of the fundamental string stretched between the two
%membranes (W boson), which shows up in the computation of the zero $p_{11}$
%transfer amplitude \cite{dos,do}. 
%It was pointed out in \cite{dos} that, for
%the gauge theory used here, the mass of the W boson
%is not equal to the geodesic distance,
%but actually is equal to the Euclidean distance in the total field space
%including the massive fields. 
%It is possible to modify the action, order by order in the normal
%coordinate expansion on ${\cal M}$, so that the classical W boson mass 
%is equal the geodesic distance \cite{dko}. However such an action is
%not renormalizable in three dimensions and cannot be used to quantize
%the membrane dynamics. 

\section{Instanton Calculation}

In this section we briefly summarize some aspects of the instanton
calculation that one needs to do in order to compute the leading
term in the membrane scattering process. We will work in the orbifold
limit where all F-I parameters vanish. In this limit we
we will see that there are precisely eight fermionic zero modes, all
originating in the broken supersymmetries.  Turning on the FI
parameters will make the calculation quite a bit more involved, but
it is clear that for sufficiently small parameters the number of
fermionic zeromodes will remain eight and that the resulting amplitude
will be nonvanishing.

To perform the instanton calculation we adapt the notation of
\cite{pp} in order to facilitate a comparison with the $N=8$ 
calculation done in that paper. Thus, we start with $N=1$
supersymmetric Yang-Mills theory in ten dimensions, and dimensionally
reduce the theory to three dimensions. This yields an $N=8$ theory
in three dimensions, which can also be viewed as an $N=4$ theory
with an extra adjoint hypermultiplet. We can then change the 
hypermultiplet representation from adjoint to some arbitrary other
representation while preserving half of the supersymmetry.

The ten-dimensional indices will be denoted by capitals
$M=1,\ldots,10$, and we will decompose this into a six dimensional
index $m=1,\ldots,6$ and an index $\alpha=7,\ldots,10$. The six
dimensional index can be further decomposed into the three-dimensional
space-time index $\mu=1,2,3$ and another index $p=4,5,6$. The
ten-dimensional gamma matrices are $\gamma_{\mu}=\sigma_{\mu} \otimes
1 \otimes \tau_1$, $\gamma_p=1\otimes \Gamma_p \otimes \tau_2$, and
$\gamma_{\alpha}=1\otimes \Gamma_{\alpha} \otimes \tau_2$. Here, $\sigma$
and $\tau$ are Pauli matrices and $\Gamma$ is a set of $8 \times 8$
seven-dimensional gamma-matrices. One finds that $\gamma_{11}=1
\otimes 1 \otimes \tau_3$ and that the charge conjugation matrix is
$C=\sigma_2 \otimes C_{(7)} \otimes \tau_1$. The ten-dimensional
fermion and the supersymmetry parameter $\epsilon$ are both
Majorana-Weyl. The reduction from ten to three dimensions is performed
by decomposing the ten-dimensional index as indicated above. The
$4,\ldots,10$ components of the gauge field become scalars in three
dimensions. $\varphi_p \equiv A_p$ are three scalar fields in the
adjoint and part of the vector multiplet, whereas $Q_{\alpha} \equiv
A_{\alpha}$ are the scalars in the hypermultiplet\footnote{Since
the hypermultiplets can be in a complex representation, one should
in fact first combine $Q_7,Q_8$ and $Q_9,Q_{10}$ into two complex
adjoint scalars and then replace the adjoint index by some other
representation}. The fermion is
decomposed in the $+$ and $-$ eigenvalues of the operator 
\be
P=\gamma_7 \gamma_8 \gamma_9 \gamma_{10}.
\ee
The $+$ eigenvalues are the adjoint fermions in the vector multiplet
denoted by $\lambda$,
whereas the $-$ eigenvalues are the fermions in the hypermultiplet
denoted by $\psi$. 
The supersymmetries that survive are those that satisfy $P\epsilon=
\epsilon$. The action of the three-dimensional $N=4$ theory reads
\be
S=\frac{1}{e^2} \int d^3 x (
\frac{1}{4} (F_{mn} F_{mn} + 2 F_{m\alpha} F_{m\alpha} + D_{\alpha\beta}
D_{\alpha\beta}) + 
\frac{i}{2} (\bar{\lambda},\bar{\psi}) \gamma_{M} D_M (\lambda,\psi) ).
\ee
Here $F_{\mu\alpha}=D_{\mu} Q_{\alpha}$, etc. The quantity
$D_{\alpha\beta}$ contains the D- and F-term equations, to turn on
a FI parameter one just has to add suitable constants to
$D_{\alpha\beta}$. 

In the orbifold limit, a BPS monopole has the property that the gauge
field $A_{\mu}^i$ is independent of $i$, i.e. only a gauge field
for the diagonal $U(2)$ is turned on, and that $Q_{\mu}^i$, the
hypermultiplet transforming as $(2_i,\bar{2}_{i+1})$, is also
independent $i$. This is precisely a normal slice as discussed in
section~3.3. The BPS equations read
\be
F_{\mu p}^i=\frac{1}{2} n_p \epsilon_{\mu\nu\rho} F_{\mu\nu}^i =
n_p B_{\mu}, \qquad
F_{\mu \alpha}^i=\frac{1}{2} n_{\alpha} \epsilon_{\mu\nu\rho} F_{\mu\nu}^i =
n_{\alpha} B_{\mu}
\ee
where $B_{\mu}=\frac{1}{2} \epsilon_{\mu\nu\rho} F_{\mu\nu}^i$ is the
magnetic field and $n_p n_p+n_{\alpha} n_{\alpha}=1$ are the
components of a unit vector. 

The fermion zero modes can be obtained from the supersymmetry
variations of $\lambda$ and $\psi$ 
\bea
\delta \lambda & =  & \frac{i}{2} F_{mn} \gamma_{m} \gamma_{n} \epsilon + 
\frac{i}{2} D_{\alpha\beta} \gamma_{\alpha} \gamma_{\beta} \epsilon \\{}
\delta \psi & = & \frac{i}{2} F_{m\alpha} \gamma_{[m} \gamma_{\alpha]}
\epsilon .
\eea
evaluated in a monopole background. This yields the zero modes
\be
\left(
\begin{array}{c} \lambda \\ \psi \end{array} \right) = 
\left( \begin{array}{c}
-B_{\mu} (\sigma_{\mu} \otimes (1+n_p\Gamma_p) \otimes 1 )\epsilon \\
-B_{\mu} (\sigma_{\mu} \otimes (n_{\alpha} \Gamma_{\alpha} ) \otimes 1 )\epsilon
\end{array}
\right) .
\ee
These are somewhat different from the ones in \cite{pp}, because the
unbroken supersymmetries do not depend on the choice of monopole.

To determine the bosonic zero modes, we need to impose a suitable
gauge. If $D_m,D_{\alpha}$ are the background covariant derivatives,
i.e. $(D_{\alpha} Q_{\beta})^{a} = (Q_{\alpha} T^a Q_{\beta})$, 
then we impose
\be
D_m a_m + D_{\alpha} q_{\alpha}=0
\ee
where $a_m$ and $q_{\alpha}$ are quantum fluctuations around $A_m$
and $Q_{\alpha}$. Then it is easy to find the three translational
zero modes indexed by $\nu=1,2,3$
\be a_m=F_{\nu m}, \qquad q_{\alpha} = F_{\nu \alpha}. \ee

The fourth zero mode is associated to $U(1)$ rotations. As in 
\cite{pp} or in \cite{swansea} one finds that it 
reads
\be
a_{\mu}=\frac{n_p F_{\mu p} + n_{\alpha} F_{\mu\alpha}}{||X-Y||}, \qquad
a_{p}=q_{\alpha}=0
\ee
The denominator is the Euclidean distance in field
space as before.

The inner products of these zero modes form a diagonal four by
four matrix, as follows from the fact that 
$\int d^3 x (F_{\mu\rho} F_{\nu\rho} + B_{\mu} B_{\nu})
 = \delta_{\mu\nu} \int B_{\rho} B_{\rho}$,
and $\int d^3 x (F_{\mu\nu} B_{\mu})=0$.

Next, we consider the one-loop determinants. The gauge fixing
term is $\sim \int d^3 x ((D_m a_m)^2 + (D_{\alpha} q_{\alpha})^2)$,
and the corresponding ghost term in the Lagrangian, $b D^2 c$,
gives rise to $\det(-D ^2)^k$ in the one-loop path integral for
the $U(2)^k$ gauge theory. The bosonic fluctuations add up to
\be
a_m \Delta_{mn} a_n + a_m \Delta_{m\alpha} q_{\alpha} + 
q_{\alpha} \Delta_{\alpha m} a_m + q_{\alpha} 
\Delta_{\alpha\beta} q_{\beta}
\ee
where $\Delta_{mn}=-D^2 \delta_{mn} - 2 F_{mn}$ and similarly for
the other components of $\Delta$. 

In order to simplify the form of the kinetic term we try to
diagonalize $F$ as much as possible. Since the quiver gauge
theory has a cyclic ${\bf Z}_k$ symmetry it is convenient to
perform a discrete Fourier transformation and replace 
$a_m^i,q_{\alpha}^i$ by $a_m^r,q_{\alpha}^r$, where $r$
is the Fourier label $r=0,\ldots,k-1$. One has to be a bit careful
as one is dealing with real fields, but the result is that the bosonic
kinetic term contains $6k$ copies of the operator $-D^2$ and 
for each $r$ one copy of the four-component operator
\be
\Delta_r=-D^2 - 2 {\rm Tr} (a^r_{\mu} [F_{\mu\nu},a^r_{\nu}]) - 2
{\rm Tr} (a^r_{\mu} [B_{\mu},\lambda_r X^r])+ 2
{\rm Tr} (\lambda_r X^r [B_{mu}, a^r_{\mu} ])
\ee
acting on the four components $(a^r_{\mu},X^r)$. Here,
we defined
\bea
X^r & = & \frac{1}{\lambda_r} \left( n_p a_p^r + 
\frac{1+\cos(2\pi r/k)}{2} n_{\alpha} q_{\alpha}^r + 
\frac{\sin(2 \pi r/k)}{2} n_{\alpha} q_{\alpha}^{-r} \right) \\{}
\lambda_r & =&  \sqrt{ |n_p|^2 + \cos^2 (r \pi/k) |n_{\alpha}|^2 } .
\eea
After a further change of basis $\Delta_r$ can be rewritten as
\be
\Delta_r = -D^2 - i(1+\lambda_r)B_{\mu} (\sigma_{\mu}\otimes 1) - 
 i(1-\lambda_r) B_{\mu} (1 \otimes \sigma_{\mu} ) .
\ee

The square of the Dirac operators that appear in the fermion kinetic
terms (one has to treat
$\psi$ and $\lambda$ together) is found to be equal to
\be \label{dirac}
-D^2 - i(\sigma_{\mu} \otimes 1 \otimes 1 + 
 n_p \sigma_{\mu} \otimes \Gamma_p \otimes 1 +
 n_{\alpha} \sigma_{\mu} \otimes \Gamma_{\alpha} \otimes 1) B_{\mu}
\ee
This acts on a combined space with $P=\pm 1$. The two linear operators
$n_p \Gamma_p$ and $n_{\alpha} \Gamma_{\alpha}$ anticommute.
Again, it is convenient to perform a discrete Fourier transformation to diagonalize
the operator (\ref{dirac}) as much as possible. The operator (\ref{dirac}) becomes block
diagonal, each block acting on $(\lambda^r,\psi^r,\lambda^{-r},\psi^{-r})$. 
The explicit form of each block is
\be
-D^2 - i\sigma_{\mu} B_{\mu} \otimes
\left(
\begin{array}{cccc}
1+|n_p| & \frac{1+\cos(\frac{2\pi r}{k})}{2} |n_{\alpha}| & 0 &
\frac{\sin(\frac{2\pi r}{k})}{2} |n_{\alpha}| \\
\frac{1+\cos(\frac{2\pi r}{k})}{2} |n_{\alpha}| & 1-|n_p| & -
\frac{\sin(\frac{2\pi r}{k})}{2} |n_{\alpha}| & 0 \\
0 & - \frac{\sin(\frac{2\pi r}{k})}{2} |n_{\alpha}| & 1+|n_p| & 
\frac{1+\cos(\frac{2\pi r}{k})}{2} |n_{\alpha}| \\
  \frac{\sin(\frac{2\pi r}{k})}{2} |n_{\alpha}| & 0 & 
\frac{1+\cos(\frac{2\pi r}{k})}{2} |n_{\alpha}| & 1-|n_p|
\end{array} \right) \otimes 1
\ee
The four by four matrix appearing here is equal to the identity
matrix plus an orthogonal matrix, and is straightforward
to diagonalize. This shows that the square of the Dirac
operator contains four copies of $\Delta_r^+$ and
four copies of $\Delta_r^-$ for each $r=0,\ldots,k-1$, where
\be \Delta_r^{\pm} = -D^2 - i(1\pm \lambda_r) B_{\mu} \sigma_{\mu}.
\ee

The total nonzero mode determinants are then
\be
\prod_{r=0}^{k-1} \frac{ \det(\Delta_r^+) 
 \det(\Delta_r^-) }{\det(\Delta_r)^{\frac{1}{2}} \det(-D^2)^2}
\ee
where $\det$ is the determinant with the omission of the zero modes.

It would be interesting to complete the one-loop calculation along the lines
of \cite{pp} and to compare to the supergravity result. We do expect 
quantitative agreement in the orbifold limit, but this clearly requires a 
rather nontrivial calculation in the quiver gauge theory. 

Finally, we will demonstrate that the instanton calculation yields a nonzero result
by showing that there are precisely eight fermionic zero modes. It is known
that $\Delta_0^+$ has two zeromodes and $\Delta_0^-$ has no
zeromodes \cite{callias}, thus we need to show that the operators 
$\Delta_r^+$ and $\Delta_r^-$ have no zeromodes for $r\neq 0$.
For this purpose consider the operator $D_{\lambda}=\sigma_{\mu} (\partial_{\mu}
+ A_{\mu}) + i\lambda \varphi$, where $A_{\mu}$ is the monopole gauge
field, $F=\ast D\varphi$ and $0\leq\lambda \leq 1$. We have
\bea
(\lambda^2 -1) \varphi^2 + D_{\lambda}^{\dagger} D_{\lambda} & = & 
-(D_{\mu} D_{\mu} + \varphi^2) -i(1+\lambda) \sigma_{\mu} B_{\mu} \\{}
(\lambda^2 -1) \varphi^2 + D_{\lambda} D_{\lambda}^{\dagger} & = & 
-(D_{\mu} D_{\mu} + \varphi^2) -i(1-\lambda) \sigma_{\mu} B_{\mu}.
\eea
The operator $ -(D_{\mu} D_{\mu} + \varphi^2)$ is the operator $-D^2$
above. We see therefore that for $\lambda<1$, $-D^2 - i(1\pm \lambda)
\sigma_{\mu} B_{\mu}$ is the sum of two positive semidefinite operators.
The only possible zeromodes are common zeromodes of $D_{\lambda}$ or
$D_{\lambda}^{\dagger}$ and $\varphi$. But such zeromodes would also
be zeromodes of $-D^2$, which has no zeromodes. Hence, $-D^2+i\lambda
\sigma_{\mu} B_{\mu}$ has no zeromodes for $0\leq\lambda<2$, and
neither have $\Delta_r^{\pm}$ for $r\neq 0$.

\section{Supergravity Computation}

For membrane scattering in flat space, the supergravity
computation for the process with
 exchange of one unit of longitudinal momentum
$\Delta p_- = 1/R$ ($R = R_s \gamma$) gives
the following coefficient multiplying the fourth
power of the transverse velocity of the membrane
\beq
  A_{flat}(b,R) \sim \left( \frac{1}{R^3b^3} 
+ \frac{3}{R^2b^4}
  + \frac{3}{Rb^5} \right)
\exp\Big(-\frac{b}{R}\Big) .
\eeq
It was shown by Polchinski and Pouliot \cite{pp} that the leading term
for $R \ll b$ agrees with the leading term in the instanton calculation
in the $N=8$ gauge theory in three dimensions. (The second 
and the third terms between the parenthesis in the right hand side
have the form of two and three-loop corrections. It would be interesting
to compare these with the gauge theory computations. In particular, the 
above form suggests that the agreement with the supergravity computation
requires a non-renormalization theorem 
beyond three-loop in the gauge theory computation.)

In the curved space case, we just have to replace the above expression by
an appropriate Green's function\footnote{This 
can be shown by treating one membrane as a source and the other as
a probe, as in \cite{pp}. Notice that because the membranes have equal 
mass this is only a good approximation for the leading term in the
velocity expansion of the scattering amplitude.} 
on $S^1 \times R^3 \times {\cal M}$. When the
distance between the two membranes
is shorter than the typical curvature radius of ${\cal M}$, we can 
use the De\thinspace Witt-Schwinger expansion to 
evaluate the Green's function for the exchange of $n$-units of $p_{11}$
as
\beq
  A_{curved}(X,Y ; R) 
   \sim \sum_{j=0}^\infty a_j(X,Y) \left( \frac{R^3}{2n^2}  
   \frac{\partial}{\partial R} \right)^j {\cal A}_{flat}
 (\sigma(X,Y), R/n),
\label{adiabatic} 
\eeq
where the coefficients $a_j$ for the Ricci-flat K\"ahler manifold is
given by
\beqa
 a_0(X,Y) &=& 1,\nonumber\\ 
 a_1(X,Y) &=&-\frac{1}{180} R_{\alpha\beta\gamma}^{~~~\mu}
 R^{\alpha\beta\gamma\nu} \partial_\mu \sigma^2(X,Y) \partial_\nu
\sigma^2(X,Y), \nonumber\\ 
 a_2(X,Y) &=& \frac{1}{180} R_{\alpha\beta\gamma\delta}
R^{\alpha\beta\gamma\delta}, ~~a_3(X,Y) = \cdots , ...
\eeqa
For $R \ll \sigma(X,Y)$, it becomes
\beqa
A_{curved}(X,Y; R) &\sim &
 \Bigg[ \frac{n^3}{R^3 \sigma^3(X,Y)} 
 + \frac{3n^2}{R^2 \sigma^4(X,Y)}
  + \frac{3n}{R \sigma^5(X,Y)} + \nonumber \\
&~~& a_1(X,Y) \Big( \frac{n^2}{2R^2 \sigma^2(X,Y)}
  - \frac{3}{2\sigma^4(X,Y)} - \frac{3R}{2n\sigma^5(X,Y)} \Big) +
\label{sugraanswer}
\\
&~~& + a_2(X,Y) \Big(\frac{n}{4R\sigma(X,Y)} + \cdots \Big) + \cdots 
 \Bigg]
\exp\Big(-n\frac{\sigma(X,Y)}{R}\Big) .\nonumber
\eeqa  

In the graviton scattering with zero $p_{11}$ transfer studied in 
\cite{dos,do}, the subleading terms in the adiabatic expansion 
(\ref{adiabatic}) were a major source of the discrepancy between the
supergravity and gauge theory computations. For finite $N$, the quantum
mechanics of D$0$ branes cannot generate such subleading terms. The
situation seems better in our case here as the subleading terms take
the form of an expansion in 
\beq
  \frac{R}{\sigma(X,Y)} \simeq \frac{e^2}{|\phi|},
\eeq
which corresponds to loop corrections in the gauge theory. As the instanton 
breaks all the supersymmetry, we expect
loop corrections to appear to all orders in the perturbative
expansion in the instanton background. 

However we have already seen that the instanton action in the gauge
theory computation is not equal to the geodesic distance $\sigma(X,Y)$ but
is smaller than that. This means that the gauge theory amplitude is
exponentially larger than the corresponding supergravity amplitude. 
The first non-trivial
curvature dependence appears in the first order correction and is
proportional to $a_1(X,Y)$. It
would be also interesting to compare this with the gauge theory
computation.

\section{Gauge Theory and Supergravity}

In this paper, we performed an instanton computation for the quiver
gauge theory, which describes the dynamics of membranes in the ALE
space in the low-energy short-distance region. We found that
the gauge theory amplitude is different from the supergravity amplitude. 

As far as we know, the quiver gauge theory is the unique theory in
three dimensions which is renormalizable and reproduces the correct
moduli space structure. In one relaxes the renormalizability
condition, it would be possible to construct a model whose instanton
action is exactly equal to $\frac{n}{e^2}\sigma(X,Y)$. The
Lagrangian constructed in \cite{dko} may well have this property 
since its low energy action is given by (\ref{effect}) with the
Euclidean distance in the last term replaced by the geodesic
distance, thereby removing the error term in (3.18). 
However, as we noted earlier, a non-renormalizable
theory cannot be used for the instanton computation if it involves 
quantities of order $\sigma(X,Y)/R_c$ or smaller, where $R_c$ is 
the typical curvature radius of
${\cal M}$. Since the issue is whether the instanton
action $S_{instanton}$ agrees with the geodesic distance
$\frac{n}{e^2}\sigma(X,Y)$, including the subleading terms in
the expansion in $\sigma(X,Y)/R_c$, the computation done using the
non-renormalizable model cannot be trusted and we have to rely on a
renormalizable gauge theory. 

What does this result imply for the large-$N$ Matrix Theory conjecture
\cite{bfss}?
%Matrix Theory is a proposal for the fundamental degrees of freedom and the 
%Hamiltonian of M Theory. It is supposed to describe Discretized Light Cone 
%Quantization (DLCQ) of M Theory \cite{suss,sei,hp}. 
%In fact it has been found that Matrix Theory
%reproduces various BPS states of M Theory and their duality relations in 
%cases with large number of supersymmetry (For review see, for example,
%\cite{banks}). 
To understand the origin of the discrepancy, it is important to examine the
region of validity of the gauge theory computation within the framework 
of the large-$N$ conjecture.
For large but finite $N$, the worldvolume of the membrane has
the short-distance cutoff given by
\beq
\delta x = \frac{L}{\sqrt{N}},
\eeq
where $L$ is the  size of the membrane \cite{bfss,ab,kt}. 
As pointed out in \cite{pp},  the length scale for the gauge theory 
on the membrane differs from that on the target spacetime by the
factor of $\gamma = Nl_p^3/R_sL^2$.
In particular the cutoff of the gauge theory is given by 
\beq
 \delta \tilde{x} = \gamma \delta x = \frac{\sqrt{N}l_p^3}{L R_s}.
\eeq
Since the size of the cutoff is crucial in the following discussion, 
we will rederive this formula in the appendix of this paper,
starting from the large-$N$ Matrix Theory.
The gauge theory description is applicable only if the cutoff
$\delta \tilde{x}$ is smaller 
than the scales set by the gauge coupling constant
\beq
 e^{-2} = \frac{l_s}{g \gamma} = \frac{L^2}{NR_s},
\eeq
and the Compton wave-length  of the W-boson 
\beq
|\phi|^{-1} = \frac{l_s^2}{\sigma(X,Y)} = \frac{l_p^3}{
  \sigma(X,Y) R_s} .
\eeq 
The condition $\delta \tilde{x} \ll e^{-2}$ requires
\beq
  l_p \ll \delta x,
\label{cutoff}
\eeq
and $\delta \tilde{x} \ll |\phi|^{-1}$ gives
\beq
  \sigma(X,Y) \ll \delta x.
\label{manycutoff}
\eeq
The gauge theory description is valid only if both
the Planck length
$l_p$ and the distance $\sigma(X,Y)$ between the membranes are
{\it shorter} than the cutoff distance $\delta x$ on the membrane. 
Thus the gauge theory and the supergravity cover complementary regimes
and there is no overlap between the two. 

The condition (\ref{cutoff}) also implies that 
the radius $R$ of the boosted M-circle is smaller than the Planck 
length because
\beq
  R = R_s \gamma = \frac{l_p^3}{\delta x^2}.
\eeq
It is interesting to note that a membrane wrapped on the 
boosted M-circle becomes a string
whose tension $\hat{l}_s^{-2}$ is equal to the cutoff scale
of the membrane as
\beq
\label{bbb}
 \hat{l}_s = \sqrt{ \frac{l_p^3}{R} }
   = \delta x. 
\eeq
Combined with (\ref{cutoff}), for
 the gauge theory description to be valid,
the length scales need to line up as
\beq
R 
\ll l_p \ll \hat{l}_s, ~~\sigma(X,Y) \ll \hat{l}_s
\label{newdkps}
\eeq  
If we identify the wrapped membrane as the fundamental IIA string,
this is exactly the condition for the D$2$-brane to be described by the
gauge theory \cite{dkps}.
Thus we have made a full circle and came back to our starting point
(1.1).  Beyond this regime, we cannot ignore
non-renormalizable interactions
generated by massive excitations of the wrapped membrane 
and there is no reason to believe
that the gauge theory correctly describes the membrane dynamics. 
The instanton computation in this paper is an explicit example of this fact.
In the gauge theory regime, the effective coupling constant is given by
\beq 
 \frac{e^2}{|\phi|} = \frac{Nl_p^3}{L^2 \sigma(X,Y)}
  = \frac{R}{\sigma(X,Y)} .
\eeq
Therefore the gauge theory is weakly coupled if $R \ll \sigma(X,Y)$.
This condition is compatible with (\ref{newdkps}). 
We conclude that, even in the large-$N$ limit
of Matrix Theory, the gauge theory description of the membrane 
is valid only in the short-distance regime. 

Another context in which the gauge theory analysis presented in
this paper may be of relevance is the scattering of gravitons in type IIB 
string theory compactified on an ALE space 
with Y-momentum transfer \cite{iib}. Here, the
starting point is M-theory compactified on a two-torus times
an ALE space. In the Seiberg-Sen limit \cite{sen,sei} we obtain
after T-dualizing the two-torus an $U(N)$ three-dimensional
quiver gauge theory. M-theory on $T^2$ is dual to type IIB string theory
on a circle, and the magnetic flux of the three-dimensional 
gauge theory corresponds to the momentum along the circle
(called Y-momentum in \cite{iib}). Thus exchange of Y-momentum
corresponds to an instanton process in the quiver gauge theory.
Repeating the analysis in this paper will once more result in a
discrepancy between the gauge theory and supergravity calculation.
In this case the resolution is probably again that there
exists a short-distance cutoff for the three-dimensional gauge theory,
so that the gauge theory only describes the short-distance regime.
%For small but finite $R_s$ massive states that
%are neglected in the three-dimensional gauge theory. 
%These include
%branes wrapping the two-cycles in the ALE space, massive
%string excitations and strings winding around a one-cycle in the
%two-torus. 
Massive states from wrapped branes give rise to an UV
cutoff in three dimensions, in a similar way as the short-distance
cutoff discussed above could be attributed to a membrane wrapped on
the boosted M-circle (see (\ref{bbb})). It would be interesting to
understand this in more detail, and to see how general this
phenomenon is.

\bigskip
\noindent
{\bf Acknowledgements}

HO would like to thank Joe Polchinski for discussions that initiated
this work. We would also like to thank Tom Banks,
Korkut Bardakci, 
Mike Douglas, Jeff Harvey, Shamit Kachru, Hitoshi Murayama,
Steve Shenker and Stefan Vandoren for discussions. 

JdB and HO would also like to thank the organizers of 
the CERN workshop, ``Non-perturbative
Aspects of Strings, Branes and Fields,'' December 8 - 12
for their hospitality. KH and HO thank Institute for Theoretical
Physics at Santa Barbara. 

This work is supported in part by NSF
grant
 PHY-951497 and by DOE grant
DE-AC03-76SF00098. 
KH and HO are also 
supported in part by NSF grant
PHY94-07194 through the Institute for Theoretical Physics. 
JdB is a fellow of the Miller Institute for Basic Research
in Science.

\appendix{Short-Distance Cutoff on Membrane}

\newcommand{\XX}{M}
It has been shown in  \cite{bss,vvk} that small fluctuations
of membranes in the large-$N$ Matrix Theory are described by 
a three-dimensional gauge theory. In order to understand
the region of validity of the gauge theory description, 
we estimate here the cutoff length of the gauge theory. 
For flat eleven-dimensional spacetime, 
$N$ D0-branes are described by the gauge theory in $(0+1)$
dimensions of \cite{halpern}. The bosonic part of the action is
\beq
S={1\over g_sl_s}\int \dd t \,\Tr\left(
(D_0 \XX^I)^2+l_s^{-4}[\XX^I,\XX^J]^2\right).
\label{QMaction}
\eeq
In the case of $S^1 \times R^6 \times {\cal M}$, we use
the quiver gauge theory of \cite{dm}.
We treat only the case with a single membrane, but the generalization to
several membranes is straightforward.

Following \cite{bfss}, we introduce canonical variables
$p$ and $q$ with the commutation relation
\beq
[q,p]={i\over N}.
\eeq
We consider the matrices $\XX^I$ as periodic functions of $p$ and
$q$ with periods $1$.
The commutator and the trace are replaced as
$[M,M']\to {i\over N}\{M,M'\}$ and
$\Tr\to N\int_{[0,1]^2}\dd p\,\dd q$,
where $\{,\}$ is the Poisson bracket.
For finite $N$, this is an approximation
which is valid only for slowly varying functions. More precisely,
the short distance cutoff is set by the uncertainty relation
\beq
\delta p\delta q\sim {1\over N}.
\label{uncer}
\eeq

A flat membrane of size $L\times L$ can be realized
\cite{bfss,dhn,bss} as the background with
\beq
\XX^1=Lp,~~~ \XX^2=Lq, ~~~\XX^3=\cdots =\XX^9=0.
\label{bckgrd}
\eeq
Here we introduce coordinates on the membrane
$x^1,x^2$ of period $L$ by $Lp=x^2, Lq=-x^1$.
With respect to these coordinates, the bracket and the trace are
expressed as
\beqa
&&[M,M']\,\longrightarrow \,
i\,{L^2\over N}\left({\partial M\over \partial x^1}
{\partial M'\over \partial x^2}-
{\partial M\over \partial x^2}
{\partial M'\over \partial x^1}\right),\\
&&~~~\Tr~~\longrightarrow ~\,{N\over L^2}
\int_{[0,L]^2}\dd x^1\,\dd x^2.
\eeqa
Now let us consider fluctuations around the background
(\ref{bckgrd}) parametrized by spatial components
of a gauge field $(A_1,A_2)$ and the transverse position $X^i$
($i=3,\ldots,9$) of the membrane as
\beqa
  \XX^1 &=& x^2 + \frac{L^2}{N} A_1 ,~~
  \XX^2 = -x^1 + \frac{L^2}{N} A_2 \nonumber \\
 \XX^i &=& X^i .
\eeqa
The covariant derivatives and commutators are given as
\beqa
D_0\XX^i &=&\partial_0 X^i - \frac{L^2}{N}
 (\partial_1 A_0 \partial_2 X^i - \partial_2 A_0 \partial_1 X^i) 
\nonumber \\
{[} \XX^r,\XX^i {]}
 &=& -i\,{L^2\over N}\partial_r X^i
 + i \left( \frac{L^2}{N} 
\right)^2 (\partial_1 A_r \partial_2 X^i - \partial_2 A_r \partial_1 X^i)
 ~~~(r=1,2) \nonumber \\
D_0\XX^r &=& \frac{L^2}{N} \partial_0 A_r
- \left(\frac{L^2}{N} \right)^2
 (\partial_1 A_0 \partial_2 A_r - \partial_2 A_0 \partial_1 A_r) 
\nonumber \\
{[}\XX^1,\XX^2{]}
&=& i\frac{L^2}{N}-i \left( {L^2\over N} \right)^2 (
\partial_1 A_2 - \partial_2 A_1) 
 + i \left( {L^2 \over N} \right)^3
 ( \partial_1 A_1 \partial_2 A_2 - 
\partial_2 A_1 \partial_1 A_2)  
\label{gau}
\eeqa
When we consider fluctuations of wavelengths 
longer than the cutoff set by
the uncertainty (\ref{uncer}), $\delta x^1\delta x^2
\sim L^2/N$, the two derivative terms in the above are
all negligible and we only have to keep the one derivative terms.
(When we compute $[\XX^1, \XX^2]^2$, the product of the
constant term and  the two derivative term in $[\XX^1,\XX^2]$ 
gives a finite contribution. However it is a total derivative
and does not contribute to the action.)
In the long wavelength regime,
 (\ref{gau}) can be rewritten in terms of the curvature as
\beq
D_0\XX^r=(L^2/N)F_{0r}, ~~~[\XX^1,\XX^2]=iL^2/N-i(L^2/N)^2 F_{12}.
\eeq
Substituting them into (\ref{QMaction}) and throwing away the constant
term, we obtain
\beqa
\lefteqn{S=}\\
&&{1\over g_sl_s}{N\over L^2}
\int\dd t\,\dd x^1\dd x^2\left[
(\partial_0X^i)^2-l_s^{-4}\left({L^2\over N}\right)^2(\partial_rX^i)^2
+\left({L^2\over N}\right)^2(F_{0r})^2-l_s^{-4}
\left({L^2\over N}\right)^4(F_{12})^2
\right]\nonumber\\
&&={l_s^3\over g_s}{N\over L^2}
\int\dd t\,\dd x^1\dd x^2\left[
(\partial_0\phi^i)^2-\left({L^2\over Nl_s^2}\right)^2
(\partial_r\phi^i)^2
+\left({L^2\over Nl_s^2}\right)^2(F_{0r})^2-
\left({L^2\over Nl_s^2}\right)^4(F_{12})^2
\right],\nonumber
\eeqa
where we have normalized the scalar field by the string tension as
$\phi^i=l_s^{-2}X^i$. The action does not look Lorentz invariant
in $(2+1)$ dimensions because of the various powers of
\beq
\gamma={Nl_s^2 \over L^2} = \frac{N l_p^3}{L^2 R_s}
\eeq
that appear in it.
In fact, these different powers
can be removed by rescaling the spatial coordinates on the membrane
\cite{pp} as
\beq
\widetilde{x}^r=\gamma x^r.
\eeq
After the rescaling, we have
\beq
S=
{l_s\over g_s\gamma}\int \dd t\,\dd \widetilde{x}^1\dd \widetilde{x}^2
\left((\partial_0\phi^i)^2-(\partial_{\tilde{r}}\phi^i)^2
+(F_{0\tilde{r}})^2-
(F_{\tilde{1}\tilde{2}})^2
\right),
\eeq
where the spatial integration is over $[0,\gamma L]\times [0,\gamma L]$.
 From this we see that the gauge coupling constant is given by
\beq
e^2={g_s\gamma\over l_s}.
\eeq
The uncertainty relation (\ref{uncer}) is translated to the coordinates
$\widetilde{x}^r$ as $\delta \widetilde{x}^1\delta\widetilde{x}^2
\sim \gamma^2L^2/N$ which 
%at best 
implies 
\beq
\delta\widetilde{x}^r\sim \gamma\,{L\over\sqrt{N}}
= \frac{\sqrt{N} l_p^3}{L R_s}.
\eeq

Since the mass of the membrane is $l_p^{-3}L^2$
the M-momentum is given by
\beq
p_{11}={N\over R_s}=(l_p^{-3}L^2){v_{11}\over\sqrt{1-v_{11}^2}},
\eeq
where $v_{11}$ is the velocity in
the eleven-th direction. Therefore,
$\gamma=Nl_p^3/R_sL^2$
can be identified as the Lorentz boost factor
$(1-v_{11}^2)^{-{1\over 2}}$ in the limit $v_{11}\sim 1$.

\end{document}